\documentclass{article}

\usepackage{eusipco2011}
\usepackage{epsfig}
\usepackage{psfrag}
\usepackage{color}
\usepackage{tabularx}
\usepackage{pstricks, pst-node, pst-plot, pst-circ}
\usepackage{moredefs}
\usepackage{algorithm,algorithmic}
\usepackage{amsmath}
\usepackage{amssymb}
\usepackage{amsfonts}

\newcolumntype{C}[1]{>{\centering\arraybackslash}p{#1}}

\newcommand{\algorithmicinput}{\textbf{input:}}
\newcommand{\INPUT}{\item[\algorithmicinput]}
\newcommand{\algorithmicoutput}{\textbf{output:}}
\newcommand{\OUTPUT}{\item[\algorithmicoutput]}

\def\PSNR{\mathrm{ PSNR}}
\def\punit{\, \mathrm}

\title{Optimized and Parallelized Processing Order for Improved Frequency Selective Signal Extrapolation}
\name{J\"urgen~Seiler and Andr\'e~Kaup}
\address{Chair of Multimedia Communications and Signal Processing, \\University of Erlangen-Nuremberg, Cauerstr. 7, 91058 Erlangen, Germany\\ \{seiler, kaup\}@LNT.de}

\begin{document}
\topmargin=0mm
\maketitle


\begin{abstract} \label{abstract}
In the recent years, multi-core processor designs have found their way into many computing devices. To exploit the capabilities of such devices in the best possible way, signal processing algorithms have to be adapted to an operation in parallel tasks. In this contribution an optimized processing order is proposed for Frequency Selective Extrapolation, a powerful signal extrapolation algorithm. Using this optimized order, the extrapolation can be carried out in parallel. The algorithm scales very good, resulting in an acceleration of a factor of up to 7.7 for an eight core computer. Additionally, the optimized processing order aims at reducing the propagation of extrapolation errors over consecutive losses. Thus, in addition to the acceleration, a visually noticeable improvement in quality of up to 0.5 dB PSNR can be achieved.
\end{abstract}


\section{Introduction} \label{sec:introduction}

Signal extrapolation is a very important task in image and video signal processing. In this process, a signal is extended from regions where the signal is known into regions where no information about the signal is available. An important application for signal extrapolation algorithms is e.\ g.\ the concealment of image distortions in the case that errors occur during the transmission of compressed image and video signals. Concealment on the one hand side is important for providing a decent quality to the viewer, even if parts of an image are distorted. On the other hand side, error concealment is necessary for reducing error propagation if a predictive coding scheme is used. In \cite{Tsekeridou2000, Kung2006} a good overview of the importance of error concealment and different error concealment techniques for images and videos can be found.

An algorithm that is suited well for such tasks is the Frequency Selective Extrapolation (FSE) from \cite{Kaup2005}. This algorithm generates a model of the signal to be extrapolated as weighted superposition of basis functions. Unfortunately, the algorithm as it is proposed up to now is not suited for a parallelized execution. But, in the recent years a trend has evolved to use multi-core designs for CPUs, DSPs or GPUs \cite{Blake2009, Karam2009}. These devices have in common that they follow a shared memory model and are able to run several threads in parallel, operating on the same data set. 

In the scope of this paper we propose an optimized processing order for FSE. This novel order has the advantage that a parallelized execution is possible, and, in addition to that, an improved extrapolation quality is achievable. Before the novel processing order is introduced in detail, FSE is briefly reviewed in the next section for providing an overview of the algorithm and for identifying the important properties that have to be taken into account for the improved and parallelizable processing order.   


\section{Frequency Selective Extrapolation}

\begin{figure}
	\centering
	\psfrag{L}[l][l][1]{$\mathcal{L}$}
	\psfrag{R}[l][l][1]{$\mathcal{R}$}
	\psfrag{A}[l][l][1]{$\mathcal{A} = \mathcal{L}\backslash \left(\mathcal{R}\cup\mathcal{B}_i\cup\mathcal{B}_o\right)$}
	\psfrag{B1}[l][l][1]{$\mathcal{B}_i$}
	\psfrag{B2}[l][l][1]{$\mathcal{B}_o$}
	\psfrag{p}[l][l][1]{$\mathfrak{b}$}
	\psfrag{m}[c][c]{$m$}
	\psfrag{n}[c][c]{$n$}
	\includegraphics[width=0.46\textwidth]{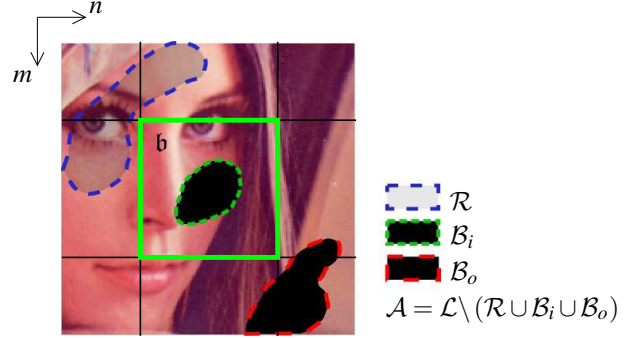}
	\caption{\emph{Extrapolation area $\mathcal{L}$ for extrapolation in \mbox{block $\mathfrak{b}$}. $\mathcal{L}$ consists of reconstructed area $\mathcal{R}$ of previously extrapolated samples, loss area $\mathcal{B}_i$ of unknown samples within block $\mathfrak{b}$, loss area $\mathcal{B}_o$ of unknown samples outside block $\mathfrak{b}$, and support area $\mathcal{A}$ of all originally known samples.}}
	\label{fig:extrapolation_area}
\end{figure}

For carrying out the Frequency Selective Extrapolation from \cite{Kaup2005}, the image to be processed is divided into blocks. The block actually being processed is denoted by $\mathfrak{b}$. If this block contains areas that have to be extrapolated, the block itself and a frame of $d$ samples width around the block are taken. This union is called extrapolation area $\mathcal{L}$ and consists of different subareas. All known samples are subsumed in support area $\mathcal{A}$, all unknown samples which are inside block $\mathfrak{b}$ are subsumed in loss area $\mathcal{B}_i$, and the unknown samples outside block $\mathfrak{b}$ in loss area $\mathcal{B}_o$. Samples that have already been extrapolated before can be used for extrapolating subsequent blocks and are subsumed in reconstructed area $\mathcal{R}$. \mbox{Figure \ref{fig:extrapolation_area}} shows the relation of the different areas with respect to the actually regarded block $\mathfrak{b}$. Altogether, extrapolation area $\mathcal{L}$ is depicted by spatial coordinates $m$ and $n$ and is of size $M\times N$.

In order to extrapolate the unknown samples, FSE iteratively generates the model 
\begin{equation}
 g\left[m,n\right] = \sum_{k\in\mathcal{K}}\hat{c}_k \varphi_k\left[m,n\right]
\end{equation}
of the signal in whole area $\mathcal{L}$ as weighted superposition of the basis functions $\varphi_k\left[m,n\right]$. Set $\mathcal{K}$ contains the indices of all basis functions used for model generation and the expansion coefficients $\hat{c}_k$ control the weight of each basis function. In every iteration, FSE selects one basis function to add to the model generated so far and estimates the corresponding expansion coefficient. Finally, the samples in area $\mathcal{B}_i$ are taken from the model and are used for extrapolation. For a detailed discussion of the original FSE and the improved expansion coefficient estimation by Orthogonality Deficiency Compensation, please refer to \cite{Kaup2005, Seiler2008}.

In order to control the influence each sample has on the model generation, FSE utilizes the weighting function 
\begin{equation}
	\label{eq:weighting_function}
	w\hspace{-0.5mm}\left[m,n\right]\hspace{-1mm}=\hspace{-1mm}\left\{\hspace{-2mm} \begin{array}{ll} \hat{\rho}^{\sqrt{\left(m-\frac{M-1}{2}\right)^2+\left(n-\frac{N-1}{2}\right)^2}} & \hspace{-3mm}\mbox{for} \left(m,n\right)\hspace{-0.5mm}\in \mathcal{A} \\ \delta\hat{\rho}^{\sqrt{\left(m-\frac{M-1}{2}\right)^2+\left(n-\frac{N-1}{2}\right)^2}} & \hspace{-3mm}\mbox{for} \left(m,n\right)\hspace{-0.5mm}\in \mathcal{R} \\ 0 & \hspace{-3mm}\mbox{for} \left(m,n\right)\hspace{-0.5mm}\in \mathcal{B}_i \hspace{-0.5mm}\cup\hspace{-0.5mm}\mathcal{B}_o\end{array}\right.
\end{equation}
within the iteration loop to assign a certain weight to each sample, depending on its position. Using $w\left[m,n\right]$, samples far away from the actual block $\mathfrak{b}$ get an exponentially decreasing weight with increasing distance, controlled by decay factor $\hat{\rho}$. As the samples within $\mathcal{B}_i$ and $\mathcal{B}_o$ are unknown, they cannot contribute to the model generation and have to be weighted with $0$. Already reconstructed samples from neighboring blocks can also contribute to the model generation. Since these samples are not as reliable as the original ones, the exponentially decreasing weight is further attenuated by a factor $\delta$ between $0$ and $1$ in area $\mathcal{R}$. This is required for reducing the influence of already reconstructed samples on the model generation and allaying error propagation.

FSE, as it is proposed up to now, extrapolates the individual blocks of an image in line scan order. This processing order has two disadvantages if consecutive losses occur. First, by just extrapolating a block without examining its neighborhood, it may be possible that blocks get extrapolated that have only few known or already reconstructed samples adjacent to them. These blocks can only be extrapolated poorly and in addition to that, the low extrapolation quality propagates to adjacent not yet extrapolated blocks.  Second, using line scan order, consecutive loss areas are processed from left to right and top to bottom. But, as already reconstructed samples are less reliable than the original ones, it is beneficial to close large consecutive loss areas from all directions in order to reduce error propagation. In order to resolve these drawbacks we subsequently propose a novel processing order for FSE. This optimized processing order is able to exploit the known samples more effectively for model generation, yielding an improved extrapolation quality. Additionally, it allows for a parallel extrapolation of different blocks, leading to an improved extrapolation speed on multi-core devices.


\section{Optimizing the Processing Order}

For optimizing the order in which the individual image blocks are processed, the properties of FSE have to be taken into account. First of all, extrapolation quality generally increases with an increasing number of known samples in the neighborhood. Due to the reuse of already extrapolated samples, this also means that a block should not be processed until as many as possible of its neighboring blocks are available. In addition to that, it is not advisable to process neighboring blocks at the same time, as the extrapolation results from one block then cannot be used for improving the model generation of the other one.

\begin{algorithm}[t]
\caption{\emph{Frequency Selective Extrapolation with optimized processing order. Function $\mathtt{neighbors}\left(\mathfrak{b}\right)$ returns indices of blocks spatially adjacent to block $\mathfrak{b}$, $\mathcal{N}$ holds the number of not extrapolated neighboring blocks for block $\mathfrak{b}$.}}
\fontsize{9}{7}\selectfont
\label{algo:proc_order}
\begin{algorithmic}
\INPUT Input signal, divided in blocks
\STATE /* Initialization */
\FORALL {blocks $\mathfrak{b}$}
	\IF {block $\mathfrak{b}$ has samples to extrapolate}
		\STATE $\mathcal{N}\left(\mathtt{neighbors}\left(\mathfrak{b}\right)\right) = \mathcal{N}\left(\mathtt{neighbors}\left(\mathfrak{b}\right)\right)+1$
	\ENDIF
	\IF {block $\mathfrak{b}$ is margin block}
		\STATE $\mathcal{N}\left(\mathfrak{b}\right) = \mathcal{N}\left(\mathfrak{b}\right)+ 3$
	\ENDIF
	\IF {block $\mathfrak{b}$ is corner block}
		\STATE $\mathcal{N}\left(\mathfrak{b}\right) =\mathcal{N}\left(\mathfrak{b}\right)+ 5$
	\ENDIF
	
\ENDFOR
\FORALL {blocks $\mathfrak{b}$}
	\IF {block $\mathfrak{b}$ has no samples to extrapolate}
		\STATE $\mathcal{N}\left(\mathfrak{b}\right) = -1$
	\ENDIF
\ENDFOR
\STATE /* Extrapolation */
\WHILE {not all blocks finished}
	\STATE $\mathcal{N}_\mathrm{min} = \min\left(\max\left(0,\mathcal{N}\left(\mathfrak{b}\right)\right)\right)$
	\STATE $\mathcal{S} = \left\{\right\}$
	\FORALL {blocks $\mathfrak{b}$}
		\IF {$\mathcal{N}\left(\mathfrak{b}\right)== \mathcal{N}_\mathrm{min} \ \ \&\& \ \ \mathtt{neighbors}\left(\mathfrak{b}\right) \notin \mathcal{S}$}
			\STATE $\mathcal{S} = \mathcal{S} \cup \mathfrak{b}$
		\ENDIF
	\ENDFOR
	\STATE /* From here on: {\bfseries parallel} execution possible */
	\FORALL {blocks $\mathfrak{b}\in \mathcal{S}$}
		\STATE Extrapolation for block $\mathfrak{b}$
		\STATE Insert extrapolated samples in block $\mathfrak{b}$
		\STATE $\mathcal{N}\left(\mathfrak{b}\right) = -1$
		\STATE $\mathcal{N}\left(\mathtt{neighbors}\left(\mathfrak{b}\right)\right) =\mathcal{N}\left(\mathtt{neighbors}\left(\mathfrak{b}\right)\right)- 1$
	\ENDFOR	
\ENDWHILE
\OUTPUT Extrapolated signal
\end{algorithmic}
\end{algorithm}

In order to obey these requirements we propose the following algorithm for determining the processing order. The main criterion for selecting a block to extrapolate is the number $\mathcal{N}\left(\mathfrak{b}\right)$ of not yet extrapolated blocks adjacent to \mbox{block $\mathfrak{b}$}. A block is regarded as not extrapolated if it contains a loss region that has not been extrapolated before, i.\ e.\ if $\mathcal{B}_i \neq \left\{\right\}$ for the regarded block. Contrary, a block is regarded as extrapolated if all samples are originally known or if the unknown samples have been extrapolated before. So, to start the extrapolation for an image, the number of not yet extrapolated neighboring blocks is determined for every block. Initially, $\mathcal{N}\left(\mathfrak{b}\right)$ is set to $0$ for all blocks. Next, the image is scanned block by block and if a block $\mathfrak{b}$ contains a loss area, $\mathcal{N}\left(\mathtt{neighbors}\left(\mathfrak{b}\right)\right)$ of all blocks adjacent to $\mathfrak{b}$ is increased \mbox{by $1$}. In doing so, function $\mathtt{neighbors}\left(\mathfrak{b}\right)$ returns the indices of the blocks spatially adjacent to block $\mathfrak{b}$. In addition to that, if a block $\mathfrak{b}$ is located at the corner of the image, $\mathcal{N}\left(\mathfrak{b}\right)$ is further increased by $5$, if a block lies at the margin of the image, $\mathcal{N}\left(\mathfrak{b}\right)$ is increased by $3$. This is due to the fact that the number of neighboring blocks is smaller for blocks that lie at the outer rim of the image. In a second blockwise scan pass, $\mathcal{N}\left(\mathfrak{b}\right)$ is set to $-1$ for all image blocks $\mathfrak{b}$ that do not contain any unknown samples. With this, all blocks for which no operation is required can be marked.

Subsequent to this, a loop is started that lasts until all image blocks that contain loss areas are extrapolated, or respectively until  $\mathcal{N}\left(\mathfrak{b}\right)<0, \forall \mathfrak{b}$. Every cycle starts with the determination of
\begin{equation}
	\mathcal{N}_\mathrm{min} = \min_\mathfrak{b}\left(\max_\mathfrak{b}\left(0,\mathcal{N}\left(\mathfrak{b}\right)\right)\right),
\end{equation}
denoting the minimal, non-negative number of not extrapolated neighbors that a block can have. Next, a list $\mathcal{S}$ is generated that holds all the blocks to be extrapolated in this run. In order to be added to this list, a block on the one hand side has to posses the just determined minimum number $\mathcal{N}_\mathrm{min}$ of not extrapolated neighbors. On the other hand side, a block is only added if none of its neighboring blocks is included in $\mathcal{S}$ yet. Following these two requirements, all the blocks of an image can be identified that posses the minimum number of not extrapolated neighbors and that can be processed independently at the same time. Thus, all blocks from list $\mathcal{S}$ can be processed in parallel and as many threads as available on the the processor can be opened for working off list $\mathcal{S}$. In this process, for every regarded block, FSE is conducted and after the model generation has been finished, the corresponding samples of the model are inserted into the loss area of the regarded block. As a processed block influences the number of not extrapolated neighbors of its adjacent blocks, $\mathcal{N}\left(\mathtt{neighbors}\left(\mathfrak{b}\right)\right)$ is decreased by $1$ in order to update the corresponding reliability information. The steps of selecting the blocks with the lowest number of not extrapolated neighbors and extrapolating them are repeated until all loss regions are filled. In order to give an overview of the proposed optimized processing order, Algorithm \ref{algo:proc_order} shows the pseudo code of the above outlined approach.

\begin{figure}
	\centering
	\includegraphics[width=0.28\textwidth]{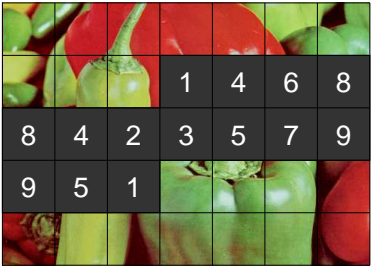}
	\caption{\emph{Example for optimized processing order. Blocks with same indices can be extrapolated in parallel. \vspace{-2.5mm}}}
	\label{fig:processing_order}
\end{figure}

Furthermore, Figure \ref{fig:processing_order} shows an example for extrapolating a large region of consecutive blocks. The numbers indicate the order in which the individual blocks are extrapolated. Blocks that posses the same number can be processed in parallel. Obviously, at the beginning, the blocks with the largest support areas $\mathcal{A}$ are selected and based on them the large consecutive loss is continuously closed. 

In general, one might also envisage different criteria for determining the processing order. These may for example take into account the number of unknown samples in a neighboring block and not only if the block contains unknown samples. But in this context, it has to be considered that even if a more precise criterion might improve the extrapolation quality a little bit more, it also harms the parallelization. This is due to the fact that less blocks exist that fulfill the more precise criterion at the same time and thus only fewer blocks can be processed in parallel.


\section{Simulation setup and results}\label{sec:results}

For evaluating the gain that can be achieved by the optimized processing order, an implementation of FSE in programming language C is regarded. The test device is a Dual quad-core AMD Opteron 2354, running at $2.2 \punit{GHz}$ and equipped with $32 \punit{GB}$ RAM. Due to the two quad-cores, the computer can effectively run up to $8$ threads in parallel. For parallelization, the shared memory base of such a computer is exploited and parallelization is conducted by using OpenMP \cite{OpenMP2010}. 

\begin{figure}
	\centering
	\includegraphics[width=0.41\textwidth]{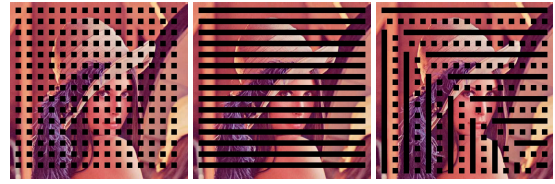}
	\caption{\emph{Examined test patterns. From left to right: Isolated losses, consecutive losses and mixed losses. \vspace{-2.5mm} }}
	\label{fig:error_pattern}
\end{figure}

Figure \ref{fig:error_pattern} shows the test patterns that are used for the evaluation. The areas to extrapolate are isolated losses, consecutive losses, and a mixture of both. The patterns consist of square areas of size $16\times 16$ samples or respectively of consecutive losses of $16$ samples width. According to \cite{Kaup2005}, the used basis functions are the ones from the Discrete Fourier Transform, as these basis functions are suited well for extrapolating smooth as well as noise like areas and edges. In addition to that, by using this basis function set an efficient implementation in the transform domain is possible as shown in \cite{Seiler2010c}. The frame of supporting samples around block $\mathfrak{b}$ is $d=16$ samples wide, the weighting function decays with $\hat{\rho}=0.8$, and already reconstructed areas are further weighted by $\delta=0.2$. The factor $\gamma$, that is necessary for Orthogonality Deficiency Compensation \cite{Seiler2008} during the model generation is set to $0.5$ and altogether $200$ iterations are carried out for generating the model.

\begin{table}[b]
 	\centering \vspace*{-5mm}
	\caption{\emph{Extrapolation quality in $\PSNR$ for test image ``Lena'' if processing is carried out in line scan order or the proposed optimized order.\vspace{2.5mm}}}
	\label{tab:lena_results}
	\begin{tabular}{|l|c|c|c|}
	 	\hline Loss pattern & Block size & Line Scan & Optimized\\
		\hline Isolated & $16\times 16$ & $25.35 \punit{dB}$ & $25.35 \punit{dB}$ \\
		\cline{2-4} & $8\times 8$ &  $25.99 \punit{dB}$ & $25.99 \punit{dB}$ \\
		\cline{2-4} & $4\times 4$ &  $25.85 \punit{dB}$ & $26.19 \punit{dB}$ \\
		\cline{2-4} & $2\times 2$ &  $25.71 \punit{dB}$ & $26.23 \punit{dB}$ \\
		\hline Consecutive & $16\times 16$ & $23.70 \punit{dB}$& $23.74 \punit{dB}$  \\
		\cline{2-4} & $8\times 8$ &  $23.83 \punit{dB}$ & $23.93 \punit{dB}$ \\
		\cline{2-4} & $4\times 4$ &  $23.74 \punit{dB}$ & $24.20 \punit{dB}$ \\
		\cline{2-4} & $2\times 2$ &  $23.47 \punit{dB}$ & $24.37 \punit{dB}$ \\
		\hline Mixed & $16\times 16$ & $24.24 \punit{dB}$ & $24.56 \punit{dB}$ \\
		\cline{2-4} & $8\times 8$ &  $24.30 \punit{dB}$ & $24.58 \punit{dB}$ \\
		\cline{2-4} & $4\times 4$ &  $24.13 \punit{dB}$ & $24.36 \punit{dB}$ \\
		\cline{2-4} & $2\times 2$ &  $24.03 \punit{dB}$ & $24.35 \punit{dB}$ \\ \hline	
	\end{tabular}
\end{table}

\begin{figure*}
	\centering
	\psfrag{error}[c][c][1]{Error pattern}
	\psfrag{linescan}[c][c][1]{Line scan order}
	\psfrag{optimized}[c][c][1]{Optimized order}
	\psfrag{24.1 dB}[c][c][1.75]{\color{white}$24.1\punit{dB}$}
	\psfrag{24.4 dB}[c][c][1.75]{\color{white}$24.4\punit{dB}$}
	\psfrag{22.9 dB}[c][c][1.75]{\color{white}$22.9\punit{dB}$}
	\psfrag{23.4 dB}[c][c][1.75]{\color{white}$23.4\punit{dB}$}
	\psfrag{18.7 dB}[c][c][1.75]{\color{white}$18.7\punit{dB}$}
	\psfrag{18.9 dB}[c][c][1.75]{\color{white}$18.9\punit{dB}$}
	\includegraphics[width=0.98\textwidth]{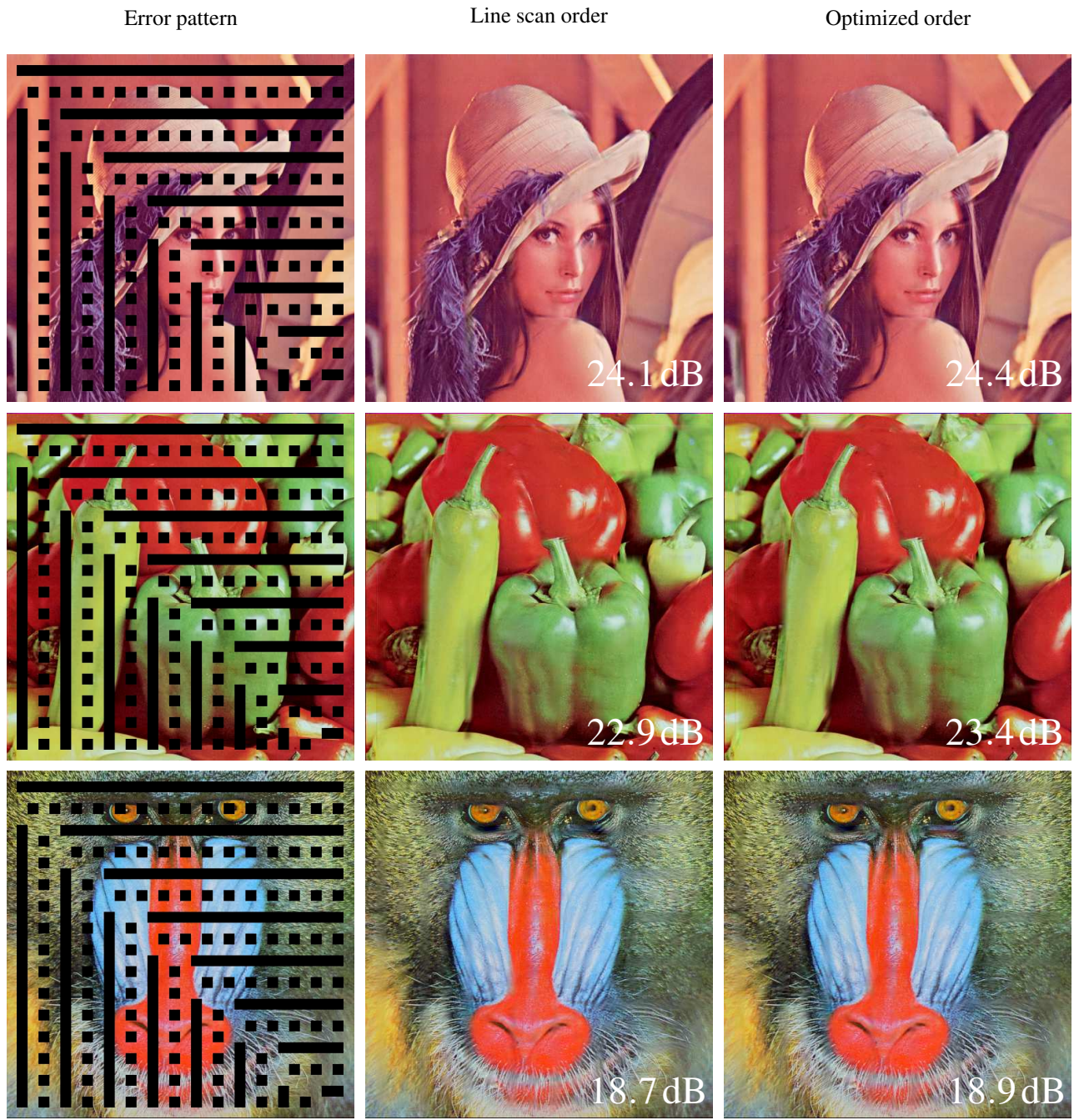}
	\caption{Visual results for the extrapolation of isolated and consecutive losses of $16$ samples width.}
	\label{fig:visual_results}
\end{figure*}

For providing some visual results, the mid and right columns of Figure \ref{fig:visual_results} show the extrapolation results for test images ``Lena'', ``Peppers'', and ``Baboon'' with the test pattern of mixed losses. The blocks to be extrapolated are set to a size of $4\times 4$ samples. Hence, the loss areas with their extent of $16\times 16$ samples are divided into $16$ small blocks for which the model is generated successively.  As the consecutive loss areas are closed from all directions instead of only from left to right, the propagation of extrapolation errors is reduced by the optimized processing order. This effect becomes especially apparent at the right side of the hat in test image ``Lena''. Comparing the Peak Signal to Noise Ratio ($\PSNR$) of the luminance component in the extrapolated areas, an improvement between $0.2\punit{dB}$ for ``Baboon'' and $0.5\punit{dB}$ for ``Peppers'' can be discovered, if the optimized processing order is used instead of line scan order. Additionally, Table \ref{tab:lena_results} shows the extrapolation quality for the considered test patterns in test image ``Lena'' with various block sizes. While the optimized processing order does not have an effect if isolated losses are concealed and the considered block size is equal to the loss size, a gain of up to $0.5\punit{dB}$ can be achieved if the area to be extrapolated is divided into smaller areas and is processed according to the proposed processing order. Furthermore, Table \ref{tab:psnr_gain} shows the average gain in $\PSNR$ for test images ``Lena'', ``Baboon'' and ``Peppers'' for the considered test patterns and different block sizes. With the only small exception of block size $8\times 8$ for the consecutive loss pattern, the optimized processing order leads to a significantly improved extrapolation quality yielding up to $0.5\punit{dB}$.

\begin{table}
 	\centering \vspace*{-2.5mm}
	\caption{\emph{Mean $\PSNR$ gain achievable by optimized processing order over line scan order.\vspace{2.5mm}}}
	\label{tab:psnr_gain}
	\begin{tabular}{|l|c|c|}
	 	\hline Loss pattern & Block size & Gain\\
		\hline Isolated & $16\times 16$ & $0.00 \punit{dB}$ \\
		\cline{2-3} & $8\times 8$ &  $0.00 \punit{dB}$ \\
		\cline{2-3} & $4\times 4$ &  $0.19 \punit{dB}$ \\
		\cline{2-3} & $2\times 2$ &  $0.33 \punit{dB}$ \\
		\hline Consecutive & $16\times 16$ & $0.00 \punit{dB}$ \\
		\cline{2-3} & $8\times 8$ &  $-0.03 \punit{dB}$\\
		\cline{2-3} & $4\times 4$ &  $0.23 \punit{dB}$ \\
		\cline{2-3} & $2\times 2$ &  $0.48 \punit{dB}$ \\
		\hline Mixed & $16\times 16$ & $0.14 \punit{dB}$\\
		\cline{2-3} & $8\times 8$ &  $0.19 \punit{dB}$ \\
		\cline{2-3} & $4\times 4$ &  $0.31 \punit{dB}$ \\
		\cline{2-3} & $2\times 2$ &  $0.44 \punit{dB}$ \\ \hline	
	\end{tabular} \vspace{-2.5mm}
\end{table}

\begin{figure}
\centering



\providelength{\AxesLineWidth}       \setlength{\AxesLineWidth}{0.5pt}
\providelength{\GridLineWidth}       \setlength{\GridLineWidth}{0.4pt}
\providelength{\GridLineDotSep}      \setlength{\GridLineDotSep}{0.4pt}
\providelength{\MinorGridLineWidth}  \setlength{\MinorGridLineWidth}{0.4pt}
\providelength{\MinorGridLineDotSep} \setlength{\MinorGridLineDotSep}{0.8pt}
\providelength{\plotwidth}           \setlength{\plotwidth}{6.5cm} 
\providelength{\LineWidth}           \setlength{\LineWidth}{0.7pt}
\providelength{\MarkerSize}          \setlength{\MarkerSize}{4pt}
\newrgbcolor{GridColor}{0.8 0.8 0.8}

\psset{xunit=0.125000\plotwidth,yunit=0.087500\plotwidth}
\begin{pspicture}(-0.718894,-0.973684)(8.092166,8.236842)


\psline[linestyle=dashed,dash=2pt 3pt,dotsep=\GridLineDotSep,linewidth=\GridLineWidth,linecolor=GridColor](0.000000,0.000000)(0.000000,8.000000)
\psline[linestyle=dashed,dash=2pt 3pt,dotsep=\GridLineDotSep,linewidth=\GridLineWidth,linecolor=GridColor](1.000000,0.000000)(1.000000,8.000000)
\psline[linestyle=dashed,dash=2pt 3pt,dotsep=\GridLineDotSep,linewidth=\GridLineWidth,linecolor=GridColor](2.000000,0.000000)(2.000000,8.000000)
\psline[linestyle=dashed,dash=2pt 3pt,dotsep=\GridLineDotSep,linewidth=\GridLineWidth,linecolor=GridColor](3.000000,0.000000)(3.000000,8.000000)
\psline[linestyle=dashed,dash=2pt 3pt,dotsep=\GridLineDotSep,linewidth=\GridLineWidth,linecolor=GridColor](4.000000,0.000000)(4.000000,8.000000)
\psline[linestyle=dashed,dash=2pt 3pt,dotsep=\GridLineDotSep,linewidth=\GridLineWidth,linecolor=GridColor](5.000000,0.000000)(5.000000,8.000000)
\psline[linestyle=dashed,dash=2pt 3pt,dotsep=\GridLineDotSep,linewidth=\GridLineWidth,linecolor=GridColor](6.000000,0.000000)(6.000000,8.000000)
\psline[linestyle=dashed,dash=2pt 3pt,dotsep=\GridLineDotSep,linewidth=\GridLineWidth,linecolor=GridColor](7.000000,0.000000)(7.000000,8.000000)
\psline[linestyle=dashed,dash=2pt 3pt,dotsep=\GridLineDotSep,linewidth=\GridLineWidth,linecolor=GridColor](8.000000,0.000000)(8.000000,8.000000)
\psline[linestyle=dashed,dash=2pt 3pt,dotsep=\GridLineDotSep,linewidth=\GridLineWidth,linecolor=GridColor](0.000000,0.000000)(8.000000,0.000000)
\psline[linestyle=dashed,dash=2pt 3pt,dotsep=\GridLineDotSep,linewidth=\GridLineWidth,linecolor=GridColor](0.000000,1.000000)(8.000000,1.000000)
\psline[linestyle=dashed,dash=2pt 3pt,dotsep=\GridLineDotSep,linewidth=\GridLineWidth,linecolor=GridColor](0.000000,2.000000)(8.000000,2.000000)
\psline[linestyle=dashed,dash=2pt 3pt,dotsep=\GridLineDotSep,linewidth=\GridLineWidth,linecolor=GridColor](0.000000,3.000000)(8.000000,3.000000)
\psline[linestyle=dashed,dash=2pt 3pt,dotsep=\GridLineDotSep,linewidth=\GridLineWidth,linecolor=GridColor](0.000000,4.000000)(8.000000,4.000000)
\psline[linestyle=dashed,dash=2pt 3pt,dotsep=\GridLineDotSep,linewidth=\GridLineWidth,linecolor=GridColor](0.000000,5.000000)(8.000000,5.000000)
\psline[linestyle=dashed,dash=2pt 3pt,dotsep=\GridLineDotSep,linewidth=\GridLineWidth,linecolor=GridColor](0.000000,6.000000)(8.000000,6.000000)
\psline[linestyle=dashed,dash=2pt 3pt,dotsep=\GridLineDotSep,linewidth=\GridLineWidth,linecolor=GridColor](0.000000,7.000000)(8.000000,7.000000)
\psline[linestyle=dashed,dash=2pt 3pt,dotsep=\GridLineDotSep,linewidth=\GridLineWidth,linecolor=GridColor](0.000000,8.000000)(8.000000,8.000000)

\psline[linewidth=\AxesLineWidth,linecolor=GridColor](0.000000,0.000000)(0.000000,0.137143)
\psline[linewidth=\AxesLineWidth,linecolor=GridColor](1.000000,0.000000)(1.000000,0.137143)
\psline[linewidth=\AxesLineWidth,linecolor=GridColor](2.000000,0.000000)(2.000000,0.137143)
\psline[linewidth=\AxesLineWidth,linecolor=GridColor](3.000000,0.000000)(3.000000,0.137143)
\psline[linewidth=\AxesLineWidth,linecolor=GridColor](4.000000,0.000000)(4.000000,0.137143)
\psline[linewidth=\AxesLineWidth,linecolor=GridColor](5.000000,0.000000)(5.000000,0.137143)
\psline[linewidth=\AxesLineWidth,linecolor=GridColor](6.000000,0.000000)(6.000000,0.137143)
\psline[linewidth=\AxesLineWidth,linecolor=GridColor](7.000000,0.000000)(7.000000,0.137143)
\psline[linewidth=\AxesLineWidth,linecolor=GridColor](8.000000,0.000000)(8.000000,0.137143)
\psline[linewidth=\AxesLineWidth,linecolor=GridColor](0.000000,0.000000)(0.096000,0.000000)
\psline[linewidth=\AxesLineWidth,linecolor=GridColor](0.000000,1.000000)(0.096000,1.000000)
\psline[linewidth=\AxesLineWidth,linecolor=GridColor](0.000000,2.000000)(0.096000,2.000000)
\psline[linewidth=\AxesLineWidth,linecolor=GridColor](0.000000,3.000000)(0.096000,3.000000)
\psline[linewidth=\AxesLineWidth,linecolor=GridColor](0.000000,4.000000)(0.096000,4.000000)
\psline[linewidth=\AxesLineWidth,linecolor=GridColor](0.000000,5.000000)(0.096000,5.000000)
\psline[linewidth=\AxesLineWidth,linecolor=GridColor](0.000000,6.000000)(0.096000,6.000000)
\psline[linewidth=\AxesLineWidth,linecolor=GridColor](0.000000,7.000000)(0.096000,7.000000)
\psline[linewidth=\AxesLineWidth,linecolor=GridColor](0.000000,8.000000)(0.096000,8.000000)

{ \small 
\rput[t](0.000000,-0.137143){$0$}
\rput[t](1.000000,-0.137143){$1$}
\rput[t](2.000000,-0.137143){$2$}
\rput[t](3.000000,-0.137143){$3$}
\rput[t](4.000000,-0.137143){$4$}
\rput[t](5.000000,-0.137143){$5$}
\rput[t](6.000000,-0.137143){$6$}
\rput[t](7.000000,-0.137143){$7$}
\rput[t](8.000000,-0.137143){$8$}
\rput[r](-0.096000,0.000000){$0$}
\rput[r](-0.096000,1.000000){$1$}
\rput[r](-0.096000,2.000000){$2$}
\rput[r](-0.096000,3.000000){$3$}
\rput[r](-0.096000,4.000000){$4$}
\rput[r](-0.096000,5.000000){$5$}
\rput[r](-0.096000,6.000000){$6$}
\rput[r](-0.096000,7.000000){$7$}
\rput[r](-0.096000,8.000000){$8$}
} 

\pspolygon[linewidth=\AxesLineWidth](0.000000,0.000000)(8.000000,0.000000)(8.000000,8.000000)(0.000000,8.000000)(0.000000,0.000000)

{  
\rput[b](4.000000,-1.3){
\begin{tabular}{c}
Threads\\
\end{tabular}
}

\rput[t]{90}(-0.8,4.000000){
\begin{tabular}{c}
Parallelization gain\\
\end{tabular}
}
} 

\newrgbcolor{color237.0034}{1  0  0}
\savedata{\mydata}[{
{1.000000,1.002521},{2.000000,2.000299},{3.000000,2.966393},{4.000000,3.917035},{5.000000,4.889183},
{6.000000,5.930501},{7.000000,6.788297},{8.000000,7.719736}
}]
\dataplot[plotstyle=line,showpoints=true,dotstyle=o,dotsize=\MarkerSize,linestyle=solid,linewidth=\LineWidth,linecolor=color237.0034]{\mydata}

\newrgbcolor{color238.0015}{0  1  0}
\savedata{\mydata}[{
{1.000000,0.999157},{2.000000,1.992047},{3.000000,2.973945},{4.000000,3.879561},{5.000000,4.762754},
{6.000000,5.848469},{7.000000,6.593554},{8.000000,7.441380}
}]
\dataplot[plotstyle=line,showpoints=true,dotstyle=+,dotsize=\MarkerSize,linestyle=solid,linewidth=\LineWidth,linecolor=color238.0015]{\mydata}

\newrgbcolor{color239.0015}{0  0  1}
\savedata{\mydata}[{
{1.000000,0.999584},{2.000000,1.968106},{3.000000,2.958779},{4.000000,3.788861},{5.000000,4.519650},
{6.000000,5.761417},{7.000000,6.242303},{8.000000,6.886634}
}]
\dataplot[plotstyle=line,showpoints=true,dotstyle=square,dotsize=\MarkerSize,linestyle=solid,linewidth=\LineWidth,linecolor=color239.0015]{\mydata}

\newrgbcolor{color240.0015}{1  0  1}
\savedata{\mydata}[{
{1.000000,0.987436},{2.000000,1.967557},{3.000000,2.936332},{4.000000,3.780123},{5.000000,4.554701},
{6.000000,5.747761},{7.000000,6.344316},{8.000000,7.181352}
}]
\dataplot[plotstyle=line,showpoints=true,dotstyle=asterisk,dotsize=\MarkerSize,linestyle=solid,linewidth=\LineWidth,linecolor=color240.0015]{\mydata}

{ 
\rput[br](7.808000,0.274286){%
\psshadowbox[framesep=0pt,linewidth=\AxesLineWidth]{\psframebox*
{\begin{tabular}{l}\small
Block size: \\
\Rnode{a1}{\hspace*{0.0ex}} \hspace*{0.5cm} \Rnode{a2}{ $2\times 2$} \\
\Rnode{a3}{\hspace*{0.0ex}} \hspace*{0.5cm} \Rnode{a4}{ $4\times 4$} \\
\Rnode{a5}{\hspace*{0.0ex}} \hspace*{0.5cm} \Rnode{a6}{ $8\times 8$} \\
\Rnode{a7}{\hspace*{0.0ex}} \hspace*{0.5cm} \Rnode{a8}{ $16\times 16$\hspace{-1mm}} \\
\end{tabular}}
\ncline[linestyle=solid,linewidth=\LineWidth,linecolor=color237.0034]{a1}{a2} \ncput{\psdot[dotstyle=o,dotsize=\MarkerSize,linecolor=color237.0034]}
\ncline[linestyle=solid,linewidth=\LineWidth,linecolor=color238.0015]{a3}{a4} \ncput{\psdot[dotstyle=+,dotsize=\MarkerSize,linecolor=color238.0015]}
\ncline[linestyle=solid,linewidth=\LineWidth,linecolor=color239.0015]{a5}{a6} \ncput{\psdot[dotstyle=square,dotsize=\MarkerSize,linecolor=color239.0015]}
\ncline[linestyle=solid,linewidth=\LineWidth,linecolor=color240.0015]{a7}{a8} \ncput{\psdot[dotstyle=asterisk,dotsize=\MarkerSize,linecolor=color240.0015]}
}%
}%
} 

\end{pspicture}%
	\caption{\emph{Parallelization gain achievable by optimized processing order for the test pattern of mixed losses. \vspace{-2.5mm}}}
	\label{fig:parallel_gain}
\end{figure}
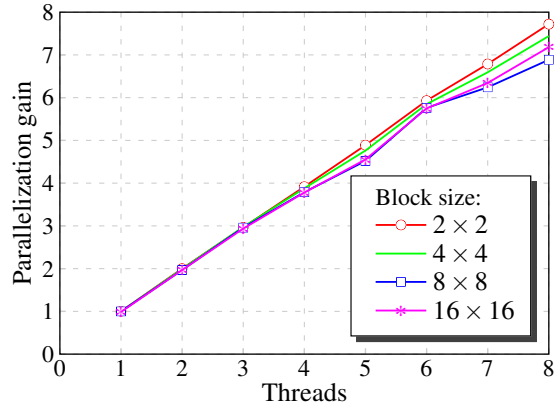

Besides the improved extrapolation quality, the gain in extrapolation speed that can be achieved by parallelization is important. To illustrate this, Figure \ref{fig:parallel_gain} shows the parallelization gain with respect to the number of used threads for different block sizes. In this context, the test pattern of mixed losses is examined. As the test platform has $8$ individual processing cores, it can execute up to $8$ threads in parallel. Looking closer at the plot, one can recognize that the algorithm scales nearly perfectly for up to $4$ threads, independently of the block size. For a larger number of threads and large block sizes, the performance drops a little bit, but the optimized processing order still is at least $6.9$ times faster than the original single thread line scan order. For small block sizes, the algorithm scales even better, achieving a parallelization gain of $7.7$. The small drop for large block sizes can be explained by the circumstance that the list $\mathcal{S}$ of blocks to be extrapolated in parallel is more often not filled completely in this case. Thus, the number of blocks that can be processed in parallel is smaller than the number of available threads. Due to this, not all threads can be fed with blocks to be extrapolated as good as for small block sizes. For larger images or larger loss areas this effect dissolves and the algorithm scales well, even for large block sizes.

\section{Conclusion} \label{sec:conclusion}

In this contribution an optimized processing order for Frequency Selective Extrapolation is introduced. This novel processing order yields an improved extrapolation quality since the area to be extrapolated is closed from all directions. In addition to that, the proposed algorithm is optimized for a parallel execution of the extrapolation in many threads and a parallelization gain of up to $7.7$ for $8$ threads is achievable. Although the algorithm has been introduced for two-dimensional signals only, it can be extended to higher dimensional signals easily by making use of \cite{Meisinger2007}. Due to the high computational complexity of the three-dimensional extrapolation, a parallelized execution that makes full use of the processing capabilities becomes even more important, there.


\end{document}